\documentstyle[12pt]{article}
\newcommand{\be}{\begin{equation}}
\newcommand{\en}{\end{equation}}
\newcommand{\bea}{\begin{eqnarray}}
\newcommand{\ena}{\end{eqnarray}}

\newcommand{\hbo}{\hbox to 1 true cm {\hfill } }
\newcommand{\tr}{\hbox{tr}}

\def\dslash{\partial\kern-.5em\slash}
\def\kslash{k\kern-.5em\slash}
\begin{document}
\vglue 1truecm

\vbox{
% pctex/gami5.t
\hfill June 14, 1994.
}

\vfil
\centerline{\bf \large Scale anomaly induced instanton
   interaction$^1$}

\bigskip
\centerline{K.\ Langfeld, H.\ Reinhardt }
\medskip
\centerline{Institut f\"ur theoretische Physik, Universit\"at
   T\"ubingen}
\centerline{D--72076 T\"ubingen, Germany }
\bigskip

\vfil
\begin{abstract}

The binary interaction of large size instantons in a SU(2) Yang-Mills theory
is obtained from the one-loop effective action for the field strength.
The instanton interaction is calculated as a function of the
instanton separation and in dependence on radius and relative orientation
of the instantons. Two equally oriented instantons with radii large
compared with the scale defined by the gluon condensate have purely
attractive interaction, whereas the interaction of maximal disoriented
instantons is repulsive. We argue that the medium range attractive
interaction of the instantons generally holds and is solely due to
the instability of the perturbative vacuum.

\end{abstract}

\vfil
\hrule width 5truecm
\vskip .2truecm
\begin{quote}
$^1$ Supported by DFG under contract Re $856/1 \, - \, 2$
\end{quote}
\eject
{\it 1.\ Introduction \/ }

The confidence in QCD as the correct theory of strong interactions
stems from the excellent agreement between QCD predictions and
high energy scattering experiments~\cite{yn83}. This success, from a
theoretical point of view, is due to QCD's remarkable property of
asymptotic freedom~\cite{yn83}, which implies that physics,
involving high momentum transfers, can be described in a perturbative
expansion with respect to the coupling.
At low energy, however, the running coupling constant is expected to be
large implying that a study of the QCD ground state is beyond the scope
of perturbation theory. For this reason only a few aspects of the
QCD vacuum are known up to now.
An important property is that scale invariance of pure
Yang-Mills theory is anomalously broken by quantum
fluctuations~\cite{co77} indicated by a non-vanishing value for the
gluon condensate.
In this context instantons~\cite{inst} play an important role
for describing the QCD vacuum, since they give rise to a gluon condensate.
Furthermore instantons may possibly trigger spontaneous breaking
of chiral symmetry~\cite{ca78} and offer an explanation of the
$U_{A}(1)$ problem~\cite{tho81}.

Instantons are gauge field configurations which minimise the
euclidean Yang Mills action and correspond to localised spots of
self- or antiself-dual field strength. Investigations of the
interaction between instantons, originating from the
classical Yang-Mills action, show that the gluonic
vacuum is not realised as a dilute gas of instantons~\cite{ca78}, but
rather than an instanton liquid~\cite{sh82,dya83,sh90}.
Recent investigations which include shape variations of the instantons
indicate that instantons might lose their identity in a strongly
correlated instanton medium~\cite{ver91}.

Instantons have a free scale parameter, the instanton radius, which
reflects the scale invariance of the classical
Yang Mills action. If the effects of fluctuations around the
instanton~\cite{tho76} are studied, a second scale provided by the
scale anomaly~\cite{co77} enters.
Calculating the one-loop effective action in dependence on the
instanton radius it was observed that the effective action of a single
instanton is
not bounded from below for large instanton radii~\cite{tho76}
indicating a infrared instability.
In order to investigate the infared behaviour of instantons in a
instanton medium the interaction of instantons
induced by quantum fluctuations is needed for large instanton radii.

In this letter we study this interaction, provided by the
effective action, of two widely separated instantons with large
instanton radii. The interaction is studied in dependence on the two
scales provided by
the instanton radius and the scale set by the trace anomaly, i.e.\
the gluon condensate. We will find
that the instanton interaction is purely attractive for instantons
with radii large compared with the scale given by the gluon condensate.
We argue that the attractive interaction at medium range
does not depend on the
details of the effective potential, but is due to the instability
of the perturbative vacuum.

\newpage
{\it 2. Scale anomaly and effective action }

The classical Yang-Mills action is invariant under scale
transformations giving rise to a conserved Noether current $j_{\mu }$ which
can be related to the trace of the energy momentum tensor~\cite{co77},
$ \partial ^{\mu } j_{\mu } = \Theta^{\mu }_{\mu } = 0 $. In the
pioneering work of Collins et al.\ \cite{co77} it was shown that the scale
invariance is broken at quantum level implying a non-vanishing
vacuum expectation value of the energy momentum tensor, i.e.
\be
\left\langle \Theta^{\mu }_{\mu } \right\rangle \; = \;
\frac{ \beta (g) }{ 2 g^{3} } \, \left\langle F^{a}_{\mu \nu }
F^{a}_{\mu \nu } \right\rangle \; ,
\label{eq:1a}
\en
where $g$ is the renormalised coupling strength,
$\beta (g) $ is the renormalisation group \break
$\beta $-function and
$F^{a}_{\mu \nu } $ is the field strength tensor. At one loop level
we have $\beta (g) \approx - \beta _{0} g^{3}$ with
$\beta _{0} = 11 N / 48 \pi ^{2}$ for a SU(N) gauge group.
Therefore (\ref{eq:1a}) implies in this case that
$F^{2}= \left\langle F^{a}_{\mu \nu } F^{a}_{\mu \nu } \right\rangle $
coincides with
$\left\langle \Theta^{\mu }_{\mu } \right\rangle $, a physical quantity,
and hence must be renormalisation group invariant
at one loop level.

The effective potential for constant chromomagnetic fields was
first calculated by Savidy~\cite{sav77}.
He found that the perturbative
vacuum is unstable, since the global minimum of the effective potential
occurs at non-zero field $H=H_{min}$. It was then pointed out by
P.\ Olesen~\cite{ole81} that the loop-expansion breaks down for fields
$H \approx H_{min}$. Furthermore it was argued~\cite{nie78} that this
breakdown for large fields $H \approx H_{min}$ is due to an instability
of the Savidy vacuum against the formation of domains of constant
magnetic field strength. For small fields the loop-expansion
is reliable. We will find (see section 4) that the medium range
instanton interaction is solely induced by the effective potential
at small field strength, where Savidy's potential can be used.

A gauge and Lorentz
covariant generalisation of the one-loop potential obtained
in~\cite{sav77}
to constant field strength $F^{a}_{\mu \nu }$ is given by
\be
V(F^{2}) \; = \; \frac{1}{4 g^{2}(\mu ) } F^{a}_{\mu \nu } F^{a}_{\mu \nu }
\; + \; \frac{ \beta _{0} }{8} F^{a}_{\mu \nu } F^{a}_{\mu \nu } \,
\ln \, F^{a}_{\mu \nu } F^{a}_{\mu \nu } / \mu ^{4} \; ,
\label{eq:1}
\en
where $\mu $ is the renormalisation point. It is easily seen that
this effective potential reflects the anomalous breaking of scale
invariance. Consider for this purpose the
scale transformation $V(F^{2}) \rightarrow V_{\lambda } =
e^{4 \lambda } V(e^{-4 \lambda }F^{2})$. One readily verifies
the correct scale anomaly at one loop level from (\ref{eq:1})
\be
\partial ^{\mu } j_{\mu } \; = \; \frac{ d }{ d \lambda } V_{\lambda }
\vert _{\lambda =0} \; = \; - \frac{ \beta _{0} }{2} F^{2} \; .
\label{eq:1b}
\en
Let us now show that the effective potential (\ref{eq:1}) follows in fact
from very general arguments. First consider the renormalisation group
equation
\be
\mu \frac{ dV}{ d \mu } \; = \;
\left( \mu  \frac{ \partial }{ \partial \mu } \; + \;
\beta (g) \frac{ \partial }{ \partial g } \right) \;
V(F^{2},\mu ,g) \; = \; 0 \; ,
\label{eq:2}
\en
where the definition $\beta (g) = \mu \frac{d g(\mu )}{d\mu }$
and the renormalisation group invariance of the field strength tensor
at one loop level,
$\frac{d }{d\mu } F^{2} =0$, was used. From dimensional arguments we have
$V(F^{2},\mu ,g) = \mu ^{4} \, f( F^{2}/\mu ^{4}, g)$, and (\ref{eq:2})
can be rewritten as a partial differential equation determining
the $F^{2}$ dependence of $V$, i.e.
\be
F^{2} \frac{ \partial V }{ \partial F^{2} } \; - \; \frac{ \beta (g) }{4}
\frac{ \partial V }{ \partial g } \; - \; V \; = \; 0 \; .
\label{eq:3}
\en
Using the boundary conditions~\cite{co77}
$\frac{ \partial V }{ \partial (1/4 g^{2}) } = F^{2}$
and $\lim _{F^{2} \to 0} V = 0$ the solution to (\ref{eq:3})
is precisely given by (\ref{eq:1}).
Renormalisation group invariance (\ref{eq:2}) also guarantees that the
effective potential $V$ does not depend on the arbitrary
subtraction point $\mu $.

For further investigations it is
convenient to rewrite $V$ in an explicitly $\mu $-independent form.
For this purpose we introduce the gluon condensate $F^{2}_{0}$
which minimises the effective potential $V(F^{2})$, i.e.
\be
\frac{ dV }{ dF^{2} } \vert _{F^{2}_{0}} \; = \;
\frac{1}{4 g^{2} } \, + \, \frac{ \beta _{0} }{8}
\left( \ln \, F^{2}_{0} / \mu ^{4} \, + \, 1 \right) \; = \; 0 \; .
\label{eq:4}
\en
Eliminating $\mu $ in the effective potential $V$ (\ref{eq:1})
in favour of $F^{2}_{0}$ with the help of (\ref{eq:4}) we obtain
\be
V(F^{2}) \; = \; \frac{ \beta _{0} }{8} F^{2} \, \left(
\ln \, \frac{ F^{2} }{ F^{2}_{0} } \, - \, 1 \right) \; .
\label{eq:5}
\en
This is the desired result because $V$ neither depend on the
renormalisation point $\mu $ nor on the renormalised coupling $g$.

In order to estimate the interaction between large size
instantons the effective action $\Gamma [ F^{a}_{\mu \nu } (x) ]$ is needed.
In leading order derivative expansion this effective action is given by
\be
\Gamma [F^{a}_{\mu \nu }(x) ]
\; = \; \int d^{4}x \; V\left( F^{2}(x) \right)
\label{eq:6}
\en
where gradients on $F^{a}_{\mu \nu }(x)$,
which occur in gauge invariant combinations of the covariant
derivative $D^{ab}_{\mu }$ and $F^{a}_{\mu \nu }(x)$, are omitted.
For an instanton medium these gradients have the order of magnitude
$1/\rho $ with $\rho $ being the instanton radius. This scale
is to be compared the with gluon condensate implying that the
terms omitted in (\ref{eq:6}) are of order $1/ F^{2}_{0} \rho ^{4}$
and are suppressed for large instantons.

We use the effective action (\ref{eq:6}) to estimate the static
interaction between instantons in SU(N) Yang-Mills theories.
We shall confine ourselves to the SU(N) instantons constructed in
reference~\cite{re93}.

\medskip
{\it 3. Binary instanton interaction }

We define the binary (anti-) instanton interaction $\Gamma _{I} $
and the interaction of an instanton anti-instanton pair,
$\Gamma _{ \bar{I} } $, by
\be
\Gamma _{I/ \bar{I} } \; := \; \Gamma [F^{(2) \, a}_{\mu \nu }(x) ] \; - \;
2 \, \Gamma [F^{(1) \, a}_{\mu \nu }(x) ] \; ,
\label{eq:15}
\en
where $F^{(2) \, k }_{\mu \nu }$ is the field strength configuration
of two (anti-) instantons or of an
instanton anti-instanton pair, respectively, and $F^{(1) \, k }_{\mu \nu }$
denote the field strengths of a single instanton.
For definiteness we will consider the binary interaction of a definite
type of instantons, obtained in ~\cite{re93}. For these instantons
the gauge potential and the corresponding field strength
is given by
\bea
A^{a}_{\mu } (x) &=& \;  G^{a}_{i} \eta ^{i}_{\mu \nu } \, x_{\nu } \,
\frac{ 2 }{ x^{2} + \rho ^{2} } \; ,
\label{eq:9a} \\
F^{a}_{\mu \nu } (x) & = & G^{a}_{i} \eta ^{i}_{\mu \nu } \,
\psi (x^{2}) \; , \hbo
\psi (x^{2}) := -  \frac{ 4 \rho ^{2} }{ (x^{2} + \rho ^{2} )^{2} } \; ,
\label{eq:9}
\ena
where $\eta ^{i}_{\mu \nu }, \, i=1 \ldots 3$ are the self-dual
't Hooft matrices and $\rho $ the instanton radius.
Anti-instantons are provided by (\ref{eq:9a},\ref{eq:9}) by replacing
$\eta ^{i}_{\mu \nu }$ by the antiself-dual 't Hooft
matrices $\bar{\eta }^{i}_{\mu \nu }$.
Furthermore, the SU(N) valued matrices $G_{i}= G^{a}_{i} t^{a} $
($t^{a}$ generators of the SU(N) group) have to fulfill an
SU(2) algebra~\cite{re93}
\be
[G_{i}, G_{k}] \, = \, i \epsilon _{ikl} \, G_{l} \; .
\label{eq:10}
\en
For $G^{a}_{i}= \delta _{ai}, \, a=1 \ldots 3$ and $G^{a}_{i}=0, \,
i = 4 \ldots N^{2}-1$ the gauge field configuration in (\ref{eq:9a})
is the SU(N) embedding of the well known SU(2) `t Hooft-Polyakov
instanton~\cite{inst}.

In order to extract long range correlations
between instantons, it is sufficient to
approximate the field strength $F^{(2) \, a }_{\mu \nu }(x)$
of two instantons separated by the
distance $r$ by the superposition of the respective field strengths
of two individual instantons, i.e.
\be
F^{(2) \, a }_{\mu \nu }(x) \; = \;
H^{a}_{k} \eta ^{k}_{\mu \nu } \, \psi (x^{2}) \; + \;
G^{a}_{i} \eta ^{i}_{\mu \nu } \, \psi \left( (x-r)^{2} \right) \; ,
\label{eq:11}
\en
where the $H_{k}=t^{a} H^{a}_{k}$ also represent an SU(2) algebra
(\ref{eq:10}). For an instanton anti-instanton configuration the
$\eta ^{k}_{\mu \nu }$ of the first term on the right hand side
of (\ref{eq:11}) is replaced by $\bar{ \eta }^{k}_{\mu \nu }$.

The effective potential only depends on $F^{2}$. For the two
instanton configuration this quantity can be
expressed as (by using $\eta ^{i}_{\mu \nu } \eta ^{k}_{\mu \nu }
= \delta _{ik}$)
\be
F^{(2) \, a }_{\mu \nu }(x) F^{(2) \, a }_{\mu \nu }(x) \; = \;
3 n_{H} \, \psi ^{2}(x^{2}) \; + \;
3 n_{G} \, \psi ^{2} \left( (x-r)^{2} \right)
\; + \; 2 \, H^{a}_{i} G^{a}_{i} \,
\psi (x^{2}) \psi \left( (x-r)^{2} \right) \; ,
\label{eq:13a}
\en
where
\be
n_{H} \; = \; \frac{1}{3} H^{a}_{i} H^{a}_{i} \; , \hbo
n_{G} \; = \; \frac{1}{3} G^{a}_{i} G^{a}_{i} \; .
\en
are the winding numbers of the two instantons. The overlap term
in (\ref{eq:13a}) depends on the relative orientations
of the two instantons $H^{a}_{i} G^{a}_{i}$. For an instanton
anti-instanton configuration this overlap term is missing
(since $\bar{\eta } ^{i}_{\mu \nu } \eta ^{k}_{\mu \nu } =0$)
and hence
\be
F^{(2) \, a }_{\mu \nu }(x) F^{(2) \, a }_{\mu \nu }(x) \; = \;
3 n_{H} \, \psi ^{2}(x^{2}) \; + \;
3 n_{G} \, \psi ^{2} \left( (x-r)^{2} \right)
\label{eq:11a}
\en
does not depend on the relative orientation of the instanton and the
anti-instanton.

For a study of the instanton instanton interaction we first consider
the simplest case of a SU(2) gauge group where the instanton (\ref{eq:9a})
coincides with the familiar instanton.
For one of these instantons, we can choose $H^{a}_{i}= \delta _{ai}$.
The $3 \times 3$ matrix $G^{a}_{i}$ then defines the orientation of the
instanton located at $x=r$ with respect to the one
at the origin and (\ref{eq:13a}) becomes
\be
F^{(2) \, k }_{\mu \nu }(x) F^{(2) \, k }_{\mu \nu }(x) \; = \;
3 \, \psi ^{2}(x^{2}) \; + \; 3 \, \psi ^{2} \left( (x-r)^{2} \right)
\; + \; 2 \, G^{k}_{k} \, \psi (x^{2}) \psi \left( (x-r)^{2} \right) \; .
\label{eq:13}
\en
The matrices $G_{k}, \, k=1 \ldots 3$ satisfying (\ref{eq:10})
can be parametrised as
$G_{k} = U t^{k} U^{\dagger } $ with the $t^{k} $ being the generators
of the SU(2) group and $U$ being an element of SU(2).
Expressing the unitary matrix $U$ by three angles $\theta ^{l}$,
$ U := \exp \{ i \theta ^{l} t^{l} \} $ the matrix $G^{k}_{i}$
takes the form of an SU(2) Wigner function
\be
G^{k}_{i} \; = \; 2 \, \tr \left( t_{k} U t_{i} U^{\dagger }  \right)
\; = \; \exp \{ - \epsilon _{kli} \theta ^{l} \} \; .
\label{eq:12}
\en
The trace of matrix $G^{k}_{i}$ entering (\ref{eq:13}) can be
straightforwardly calculated, i.e.
\be
\kappa ^{(2)} \; := \; G^{k}_{k} \; = \; 1 \, + \, 2 \, \cos
\left( \sqrt{ \theta ^{k} \theta ^{k} } \right)
\label{eq:14}
\en
implying that the effective action of the field strength for two
fixed instantons only depends on $\theta ^{k} \theta ^{k} $ in the case
of SU(2).
Note that due to our ansatz (\ref{eq:11})
the interaction of an instanton anti-instanton pair $\Gamma _{\bar{I}}$
is related to that of two instantons $\Gamma _{I}$ by
$ \Gamma _{\bar{I}} = \Gamma _{I}( \kappa =0 )$.

In the case of an SU(3) gauge group an instanton which is not
an trivial embedding of the 't Hooft Polykov instanton, but has non-trivial
colour and Lorentz structure was given in~\cite{re93}.
It has winding number $n_{G}=4$ and the corresponding matrices
$G_{i}$ in (\ref{eq:10}) form the spin one representation.
Its colour components in (\ref{eq:9a}) are
\be
G^{7}_{1} \; = \; 2 \; , \hbo G^{5}_{2} \; = \;- 2 \; , \hbo
G^{2}_{3} \; = \; 2 \; .
\label{eq:20}
\en
In the following we study their interaction induced by the effective
action (\ref{eq:6}).

The orientation $H^{a}_{i}$ of the second instanton is described
analogously to the SU(2) case
by eight angles $\theta ^{a} $ defining the gauge rotation in SU(3), i.e.
\be
H^{a}_{i} \; = \; 2 \; \tr \{ U t^{a} U^{\dagger } t^{b} \} \,
G^{b}_{i} \; , \hbo U \; = \; \exp \{ i \theta ^{a} t^{a} \} \; .
\label{eq:21}
\en
In order to calculate the coefficient (relative orientation)
$\kappa ^{(3)}:=H^{a}_{i} G^{a}_{i}$
of the interference term in (\ref{eq:13a}) we exploit
$$
\sum_{l} (G_{l})_{rs} (G_{l})_{ik} \; = \;
\sum _{a=2,5,7} t^{a}_{rs} t^{a}_{ik} \; = \; \frac{1}{4} \left(
\delta _{rk} \delta _{is} \, - \, \delta _{ri} \delta _{sk}
\right)
$$
which holds since the $G_{l}$ span here the spin one representation.
We therefore obtain
\bea
\kappa ^{(3)} & = & 2 \, \tr U \, \tr U^{\dagger } \; - \;
2 \, \tr U U^{*}
\label{eq:23} \\
& = & 4 \left[
\cos ( \lambda _{1} - \lambda _{2} ) + \cos ( \lambda _{1} - \lambda _{3} )
+ \cos ( \lambda _{2} - \lambda _{3} ) \right] \; ,
\nonumber
\ena
which is expressed in terms of the three eigenvalues $\lambda _{i}$
of the hermitean matrix $\theta ^{a} t^{a}$.

The instanton interaction $\Gamma _{I/ \bar{I}}$
as defined in (\ref{eq:15}) was investigated
numerically. Figure 1 shows $\Gamma _{I/ \bar{I}}$ as
function of the
instanton distance $r$ for various instanton sizes and
for uniquely oriented instantons ($\theta ^{k}=0$
in (\ref{eq:14})).
For all instanton radii $\rho $ the interaction is medium range attractive.
For instanton radii small compared with scale set by the gluon condensate
the interaction becomes short range repulsive. For large size instantons
the binary interaction of instantons with same orientation is
purely attractive. One should note, however, that due to the use of
(\ref{eq:11}) the estimate of the instanton interaction becomes
unreliable for instanton separations as small as the instanton radius.

In the case of SU(2) Figure 2 shows $\Gamma _{I}$ for several
instanton orientations $\kappa ^{(2)}$, which range from $-1$ to $3$.
The lowest
action is obtained for uniquely oriented instantons ($\kappa ^{(2)}= 3$),
whereas for maximal disoriented ones ($\kappa ^{(2)} = -1$) the instanton
interaction becomes purely repulsive. The same is also true for
the instanton anti-instanton interaction ($\kappa =0$).

\medskip
{\it 4.\ Discussions and conclusions }

The exact effective potential of  $F^{2}$ is mainly determined
by two phenomenological ingredients.
First, anomalous breaking of scale invariance occurs, which
implies that the effective potential is minimal for a non-vanishing
value of $F^{2}$. Second, the perturbative vacuum
is unstable implying that at $F^{2}=0$ the effective potential is a
decreasing function of $F^{2}$.
Therefore the effective potential is expected to have
the qualitative behaviour described by the potential $V$ in (\ref{eq:5}).
Let us now assume that instantons are the dominant field configurations
contributing to the QCD functional integral. In this
case the ensemble of interacting instantons should produce
an effective potential which qualitatively behaves like $V$ in
(\ref{eq:5}).
With this assumption we may interpret the binary instanton interaction
studied in the present letter as the interaction of two instantons moving
in the mean field of the instanton ensemble.

Most striking feature
of the interaction of two equally oriented instantons is a
medium range attractive force. We argue that it is solely due to the
instability of the perturbative vacuum. First note that for
widely separated instantons (with field strength $F_{(1)}$ and
$F_{(2)}$) the main contribution to the integral
$\int d^{4}x \; V \left( (F_{(1)}+F_{(2)})^{2} \right) $
stems from the space-time region
near the centers of the instantons. Since for widely separated
instantons the field strength of the first instanton is small
at the center of the second instanton, we may expand the above
integral and obtain
\bea
\Gamma _{I} & \approx & 2 \int _{(1)} d^{4}x \,
V'\left( F^{2}_{(1)} \right) \, F_{(1)} \cdot F_{(2)} \; + \;
\int _{(2)} d^{4}x \,
V'\left( F^{2}_{(2)} \right) \, F_{(2)} \cdot F_{(1)}
\nonumber \\
& = & 4 \int _{(1)} d^{4}x \,
V'\left( F^{2}_{(1)} \right) \, F_{(1)} \cdot F_{(2)} \; ,
\label{eq:24}
\ena
where $V'(F^{2}):= \frac{ d }{ dF^{2} }V (F^{2})$.
Thereby the space time region $(1)$, where $F_{(1)}^{2}$ has its
maximum, is separated from the region $(2)$ (containing the center
of the instanton $(2)$) by the hyperplane defined by
$F_{(1)}^{2}(x) = F_{(2)}^{2}(x)$. Figure 3 compares the
approximation (\ref{eq:24}) with the exact numerical result for the
effective potential $\Gamma _{I}$. The asymptotic behaviour $(r \rightarrow
\infty )$ of the effective potential is correctly described
by (\ref{eq:24}).
Since $F^{2}_{(2)}$ is a smooth function in the space time region $(1)$,
the dominant contribution in (\ref{eq:24}) stems from the region where
either $F_{(1)}$ or $V'$ is large. At the center of instanton $(1)$
the integrand is $ V'(F_{(1)}^{2}(0)) \, F_{(1)}(0) \cdot F_{(1)}(r) $
and therefore of order ${\cal O} \left( F_{(1)}(r) \right)$.
On the other hand $V'(F^{2})$ diverges for small $F^{2}$
as suggested by the perturbation theory.
Note that perturbation theory, based on a trivial vacuum, is expected
to yield reliable results for the effective potential at small $F^{2}$.
We therefore can rely on the result (\ref{eq:5}) at small $F^{2}$
implying $V'(F^{2}) = \ln F^{2}/ F^{2}_{0} $.
At the hyperplane the integrand in (\ref{eq:24})
is therefore of order $ {\cal O} \left( \ln [F_{(1)}^{2}(r)] \;
F_{(1)}(r) \right) $.
This implies that behaviour of $V'(F^{2})$ in (\ref{eq:24})
near the perturbative vacuum ($F^{2}=0$) is relevant, where
we have $V'<0$ due to the instability of the trivial
vacuum. Note further that for equally oriented instantons
$F_{(1)} \cdot F_{(2)}$ is positive, and we end up with a medium range
attractive instanton instanton interaction.

In conclusion, we have investigated the binary interaction of
instantons with size large compared with the scale provided by
the gluon condensate. This interaction is derived from the effective
action for $F^{2}$. The effective potential at small field strength, which
is accurately estimated by the leading order of the loop expansion,
is relevant for the instanton interaction at medium range.
In order to estimate the short range interaction of instantons, the
effective potential is also required for large fields, which is
beyond the validity of the loop expansion. For large fields we
modelled an effective potential consistent with renormalisation
group arguments and phenomenological requirements.

We find that the interaction of two uniquely oriented instantons
with radii large compared with the gluon condensate is purely
attractive. Equally oriented instantons have minimal effective action.
Since in this case no repulsive core prevents the large size instantons
from collapsing, this result might indicate a condensation of
large scale instantons forming a homogeneous gluon condensate.
The instanton anti-instanton interaction induced by the
scale anomaly is repulsive for all instanton orientations.
We note, that our results
for the short range instanton interaction
depend on our model assumptions of the effective potential at large
field strength and are strictly valid only for large instantons
separations.

The situation can be compared with that of an Heisenberg
spin system, where
ferromagnetism is obtained by investigating the binary interaction
of spins induced by the mean magnetisation of the spin ensemble.
In analogy, it might happen that
the instanton interaction accounts for a gluon vacuum which
decomposes into domains of a coherent superposition of (anti-) instantons.
This picture of the gluonic ground state is consistent with that
emerging in the field strength approach to Yang-Mills
theories~\cite{sch90,la92a}.
The precise form of the instanton vacuum of QCD will likely be
determined by the interplay between the attractive instanton interaction
(favouring a homogeneous instanton condensate) and the entropy
(favouring a domain or liquid type vacuum).
This is an challenging subject for future work which requires
a detailed numerical simulation of the instanton
medium~\cite{sh82,dya83,sh90}
using the above extracted instanton interaction as ingredient.

\newpage
\centerline{ \bf \large Figure captions: }

\vspace{2cm}
{\bf Figure 1: }
   The interaction $\Gamma _{I}$ of two uniquely oriented instantons
   as function of the
   instanton distance $r$ for several instanton radii $\rho $.
   $r$ and $\rho $ in units of $( F^{2}_{0} )^{-1/4}$.
   $\Gamma _{I}$ in units of the gluon condensate $F_{0}^{2}$.

\bigskip
{\bf Figure 2: }
   The (anti-) instanton interaction $\Gamma _{I}$ as function of the
   instanton distance $r $ for several instanton orientations $\kappa $
   and for $\rho = 2$.
   $r$ and $\rho $ are given in units of $( F^{2}_{0} )^{-1/4}$.
   $\Gamma _{I}$ in units of the gluon condensate $F_{0}^{2}$.
   The interaction of an instanton anti-instanton pair $\Gamma _{\bar{I}}$
   is also shown ($\kappa =0 $).

\bigskip
{\bf Figure 3: }
   The instanton interaction $\Gamma _{I}$ provided by the
   approximation (23) compared with the exact numerical
   result for $\kappa =3$ and $\rho = 2$.
   $r$ and $\rho $ are given in units of $( F^{2}_{0} )^{-1/4}$.
   $\Gamma _{I}$ in units of the gluon condensate $F_{0}^{2}$.


\begin{thebibliography}{sch90}
\bibitem{yn83}{ see e.g. F.\ J.\ Yndurain,
   'Quantum Chromodynamics', Springer Verlag, 1983.}
\bibitem{co77}{ J.\ C.\ Collins, A.\ Duncan, S.\ D.\ Joglekar,
   Phys.Rev. D16(1977)438. }
\bibitem{inst}{ A.\ M.\ Ployakov, Phys.Lett. B59(1975)82. \\
   A.\ A.\ Belavin, A.\ M.\ Polyakov, A.\ A.\ Schwartz,
   Yu.\ S.\ Tyupkin, \\ Phys.Lett. B59(1975)85. \\
   A.\ A.\ Belavin, A.\ M.\ Polyakov, JETP Lett 22(1975)245. \\
   G.\ 't Hooft, Phys.Rev.Lett. 37(1976)8. }
\bibitem{ca78}{ C.\ G.\ Callan, R.\ Dashen, D.\ J.\ Gross, Phys.Rev.
   D17(1978)2717. \\
   C.\ G.\ Callan, R.\ Dashen, D.\ J.\ Gross, Phys.Rev.
   D19(1978)1826. }
\bibitem{tho81}{ G.\ 't Hooft, Nucl.Phys. B190(1981)455. \\
   G.\ 't Hooft, Phys.Scr. 25(1981)133. }
\bibitem{sh82}{ E.\ V.\ Shuryak, Nucl.Phys. B203(1982)93. }
\bibitem{dya83}{D.\ I.\ Dyakonov, V.\ Yu.\ Petrov,
   Phys.Lett. B130(1983)385.  \\
   D.\ I.\ Dyakonov, V.\ Yu.\ Petrov, Nucl.Phys. B245(1984)259. }
\bibitem{sh90}{ E.\ V.\ Shuryak, J.\ J.\ M.\ Verbaarschot, Nucl.Phys.
   B341(1990)1. }
\bibitem{ver91}{ J.\ J.\ M.\ Verbaarschot, Nucl.Phys. B262(1991)33. }
\bibitem{tho76}{ G.\ 't Hooft, Phys.Rev. D14(1976)3432. \\
   C.\ Bernard, Phys.Rev. D19(1979)3013. }
\bibitem{sav77}{ G.\ K.\ Savidy, Phys.Lett. B71(1977)133. \\
   S.\ G.\ Matinyan, G.\ K.\ Savidy, Phys.Lett. B134(1978)539. }
\bibitem{ole81}{ P.\ Olesen, Physica Scripta 23(1981)1000. }
\bibitem{nie78}{ N.\ K.\ Nielsen, P.\ Olesen, Nucl.Phys. B144(1978)376. }
%\bibitem{at77}{ M.\ F.\ Atiyah, R.\ S.\ Ward, Comm.Math.Phys. 55(1977)177. \\
%   M.\ F.\ Atiyah, N.\ J.\ Hitchin, V.\ G.\ Drinfeld, Yu.\ I.\ Manin, \\
%   Phys.Lett. A65(1978)185. }
%\bibitem{co78}{ E.\ F.\ Corrigan, D.\ B.\ Fairlie, S.\ Templeton,
%   P.\ Goddard, \\ Nucl.Phys. B140(1978)31. }
\bibitem{re93}{ H.\ Reinhardt, K.\ Langfeld, Phys.Lett. B317(1993)590. }
\bibitem{sch90}{ M.\ Schaden, H.\ Reinhardt, P.A.\ Amundsen, M.J.\ Lavelle \\
  Nucl.Phys. B339(1990)595. \\
  P.A.\ Amundsen, M.\ Schaden, Phys.Lett. B252(1990)265. }
\bibitem{la92a}{ K.\ Langfeld, H.\ Reinhardt, 'Instanton condensation
   field strength formulated QCD',  UNIT"U-THEP-18$/$1992,
   hep-ph 9301230. }


\end{thebibliography}
\end{document}